\documentclass[aps,prb,showpacs,twocolumn,superscriptaddress]{revtex4}

\usepackage{amssymb,amsmath,amsfonts,amsthm}
\usepackage{mathrsfs}
\usepackage[mathcal]{eucal}

\usepackage[dvips]{graphicx}

\input{epsf}

\renewcommand{\Im}{\mathop{\rm Im}\nolimits}
\renewcommand{\Re}{\mathop{\rm Re}\nolimits}

\newcommand{\ds}{\displaystyle}
\newcommand{\uu}{\uparrow\!\uparrow}
\newcommand{\dd}{\downarrow\!\downarrow}
\newcommand{\ud}{\uparrow\!\downarrow}
\newcommand{\du}{\downarrow\!\uparrow}
\newcommand{\vsigma}{\mbox{\boldmath $\sigma$}}

\newcommand{\vpi}{\mbox{\boldmath $\pi$}}

\newcommand{\Tr}{\mathop{\rm Tr}\nolimits}
\newcommand{\Sp}{\mathop{\rm Sp}\nolimits}
\newcommand{\tr}{\mathop{\rm tr}\nolimits}

\begin{document}

\title{Diffusive magnetotransport in a two--dimensional Rashba system}

\author
{S.G.~Novokshonov}
\affiliation
{Institute of Metal Physics, Ural Division of RAS, Ekaterinburg,
Russia}
\author
{A.G.~Groshev}
\affiliation
{Physical-Technical Institute, Ural Division of RAS, Izhevsk, Russia}

\date{\today}

\begin{abstract}
An analytical approach to calculation of the conductivity  tensor,
$\sigma$, of a two dimensional (2D) electron system with Rashba
spin--orbit interaction (SOI) in an orthogonal magnetic field is
proposed. The electron momentum relaxation is assumed to be due to
electron scattering by a random field of short--range impurities,
which is taken into account in the Born approximation. An exact
expression for the one--particle Green function of an electron
with Rasba SOI in an arbitrary magnetic field is suggested. This
expression allows us to obtain analytical formulas for the density
of states (DOS) and $\sigma$ in the self--consistent Born and
ladder approximations, respectively, which hold true in a wide
range of magnetic fields, from the weak $(\omega_{c}^{}\tau\ll 1)$
up to the quantizing $(\omega_{c}^{}\tau\gtrsim 1)$ ones. It is
shown that in the ladder approximation the Rashba SOI has no
effect at all on the conductivity magnitude in the whole range of
classical (non quantizing) magnetic fields. The Shubnikov--de Haas
(SdH) oscillation period is shown to be related to the total
charge carrier concentration by the conventional formula,
irrespective of the SOI magnitude. A simple equation defining the
location of the SdH oscillation beating nodes is obtained. The
results are in good agreement whith the experimental and recent
numerical investigations.

\end{abstract}

\pacs{
71.70.Ej,  
72.15.Gd,  
73.20.At,  
73.21.-b}  

\maketitle

\section{Introduction}

The growing interest in studying the spin--orbit interaction (SOI) in
semiconductor two dimensional (2D) structures is mostly due to its potential
application to the spin--based electronic devices \cite{zutic_etal_2004}. There
are two main types of SOI in the quantum well based on zinc--blende--lattice
semiconductors. First, the Dresselhaus interaction \cite{dressel_1955} that
originates from the bulk inversion asymmetry (BIA); second, the Rashba
interaction \cite{rashba} induced by structural inversion asymmetry (SIA) of
the confining field of a quantum well. Both of these interactions lead to the
momentum-dependent spin splitting of the electron energy spectrum and to the
formation of quantum states with the hard linked spatial and spin degrees
of freedom of the electrons. They are responsible for many interesting effects
in the transport phenomena like beatings in the Shubnikov--de Haas (SdH)
oscillations \cite{rashba,luo_etal}; weak antilocalization
\cite{iord_etal_1994,pikus_pikus_1995,knap_etal_1996,golub_2005};
current--induced non--equilibrium spin polarization
\cite{levit_etal_1985,edelst_1990}; spin Hall effect
\cite{dyak_etal_1971,mura_etal_2003}, and so on.

At present there are some sufficiently well developed theories of the kinetic
and spin phenomena in 2D systems with SOI in zero or classical weak
$(\omega_{c}^{}\tau\ll 1)$ orthogonal magnetic fields. Here $\omega_{c}^{}=
|e|B/mc$ is the cyclotron frequency, and $\tau$ is the electron scattering time.
As for theoretical studies of the considered systems in strong, and especially
in quantizing $(\omega_{c}^{}\tau\gtrsim 1)$ magnetic fields, there is still no
satisfactory analytical description of the kinetic phenomena even in the
usual diffusive regime (without quantum corrections). The complex form of the
eigenspinors and energy spectrum of an electron in the presence of SOI and
a strong magnetic field \cite{rashba} is the main cause of such a situation.
Direct employment of this basis forces one to proceed almost right
from the start to the numerical analysis of very cumbersome expressions
\cite{wang_etal_2003,lange_etal_2004,averk_etal_2005,wang_etal_2005}.

The strong magnetic field is however one of the most efficient tools for
investigation of SOI \cite{luo_etal} and manipulation of the spin degrees
of freedom in the semiconductor 2D--structures. Thus, a rather simple
theoretical description of the kinetic phenomena in the 2D systems with
SOI in a strong orthogonal magnetic field becomes a necessity. In the present
work, we consider the problem of calculation of the longitudinal and Hall
resistances of a 2D Rashba system in the ladder approximation assuming
that the electron momentum relaxation is due to elastic scattering by
short--range impurities which is taken into account in the Born
approximation.

We have found the {\it exact} relation between the one--particle
Green function (GF) of the Rashba 2D--electron in an arbitrary
orthogonal magnetic field and the well known GF of an "ideal"\,
electron, that is an electron with the ideal value of the Zeeman
coupling $(g_{0}^{}=2)$ and without SOI. This allows one to obtain
the analytical expressions for the density of states (DOS) in the
self--consistent Born approximation (SCBA), and the conductivity tensor
$\hat\sigma$ in the ladder approximation. The total DOS in SCBA is defined
as the sum of the partial DOS's of two spin--split subbands. At the same time,
conductivity in ladder approximation looks as if the current were generated by
charge carriers of one type with the total concentration $n$ and mobility $\mu$. 
These expressions hold good in a wide range, from the
classically weak magnetic fields $(\omega_{c}^{}\tau\ll 1)$ up to the quantizing
ones $(\omega_{c}^{}\tau \gtrsim 1)$.
On the basis of these results, we perform the numerical analysis of
the beatings of the SdH oscillations of the considered kinetic coefficients,
as well as of their behavior in the classical magnetic fields region.
The results are in good agreement with the experimental data
\cite{nitta_etal,das_etal}, and with the results of a recent numerical
investigation \cite{yang_chang_2006}.

\section{Model}

Let us consider a two--dimensional $(|| OXY)$ degenerate gas of electrons with
effective mass $m$, and effective Zeeman coupling $g$ those move in an external
orthogonal $({\bf B}|| OZ)$ magnetic field ${\bf B}=\nabla\times{\bf A}$ in the
presence of a random field $U({\bf r})$ due to pointlike impurities distributed
by Poisson law in the sample. We assume the Rashba interaction to be the
dominant mechanism of the energy spin splitting in the absence of a magnetic
field. This situation occurs, for example, in the narrow--gap semiconductor
heterostructures, such as ${\rm In\,As}/{\rm Ga\,Sb}$ \cite{luo_etal},
${\rm In\,Ga\,As}/{\rm In\,Al\,As}$ \cite{das_etal,nitta_etal}. The
one--particle Hamiltonian of the considered system has the form
\begin{equation}
\label{eq:model_h}
{\cal H}+U=\frac{\vpi_{}^{2}}{2m}+\alpha(\vsigma\times\vpi)\cdot{\bf n}
+\frac{1}{4}g\omega_{c}^{}\sigma_{z}^{}+U({\bf r})
\end{equation}
($\hbar=1$). Here $\vpi={\bf p}-e{\bf A}/c=m{\bf v}$ is the operator of the
kinematic electron momentum; $\vsigma=(\sigma_{x}^{},\,\sigma_{y}^{},\,
\sigma_{z}^{})$ is the vector formed by the Pauli spin matrices; $\alpha$ is
the Rashba spin--orbit coupling; $g$ is the effective Zeeman coupling
($g$--factor).

In the gauge ${\bf A}=(0,Bx,0)$, the components of the
eigenspinors of the Hamiltonian ${\cal H}$ (\ref{eq:model_h}) of a
free ($U({\bf r})=0$) Rashba electron are expressed through the
Landau wave functions $\psi_{n,X}^{} ({\bf r})$ depending on the
Landau level number $n=0,1,2,\ldots$ and the $X$--coordinate of
the cyclotron orbit center $X=-k_{y}^{}/m\omega_{c}^{}$
\cite{rashba}
\begin{subequations}
\label{eq:r_basis}
\begin{equation}
\label{eq:r_spinor}
\ds\widehat\Psi_{\alpha}^{}({\bf r})=\frac{1}
{\sqrt{1+C_{s,n}^{2}}}\left[\begin{array}{c}
\ds C_{s,n}^{}\psi_{n-1,X}({\bf r})\\[4pt]
\ds\psi_{n,X}^{}({\bf r})\end{array}\right]\,,\quad\alpha=(s,n,X)\,.
\end{equation}
The corresponding energy levels have the following form
\begin{equation}
\label{eq:r_spectrum}
{\cal E}_{s,n}^{}=\left\{\begin{array}{ll}
-\omega_{c}\delta\,,  & n=0,~~s=+1\,,\\
\omega_{c}^{}\big[n+s\sqrt{\delta_{}^{2}+2\gamma_{}^{2}n}\,\big]\,, &
n>0,~~s=\pm 1\,.\end{array}\right.
\end{equation}
\end{subequations}
Here $C_{s,n}^{}=\gamma\sqrt{2n}/\big[s\sqrt{\delta_{}^{2}+2\gamma_{}^{2}n}
-\delta\big]$ is a normalizing coefficient; $\delta=(g-2)/4$ is the relative
deviation of the effective Zeeman coupling from its ideal value $g_{0}^{}=2$
(for definiteness, it is assumed that $\delta<0$ in these equations, but all
the following results are valid for any sign of $\delta$); and, finally, $\gamma
=\alpha\sqrt{m/\omega_{c}^{}}$ is the dimensionless Rashba spin--orbit coupling.

The quantum number $s=\pm 1$ describes the {\it helicity} of the Rashba
electron eigenstate as in the absence of a magnetic field \cite{edelst_1990}.
Indeed, it can be verified immediately that $s=\pm 1$ is the eigenvalue of the
operator
\begin{equation}
\label{eq:helic_oper}
\nu=\frac{\big[\alpha\vsigma\times\vpi+\omega_{c}^{}\delta\vsigma\big]\cdot
{\bf n}}{\sqrt{2m\alpha_{}^{2}{\cal H}_{0}^{}+
\omega_{c}^{2}\delta_{}^{2}}}\,,
\end{equation}
that is diagonal in the basis (\ref{eq:r_spinor}) and approaches the helicity
operator $(\vsigma\times{\bf p})\cdot{\bf n}/|{\bf p}|$ as $B\to 0$.
Here ${\bf n}$ is the unit normal vector to the considered 2D system;
\begin{equation}
\label{eq:ideal_h}
{\cal H}_{0}^{}=\frac{\vpi_{}^{2}}{2m}+\frac{1}{2}\omega_{c}^{}\sigma_{z}^{}
\end{equation}
is the Hamiltonian of the "ideal"\, ($g_{0}^{}=2$) electron in a magnetic field,
which commutes with $\vsigma\cdot{\bf n}$, $(\vsigma\times\vpi)\cdot{\bf n}$,
and with ${\cal H}$ (\ref{eq:model_h}).

In spite of this analogy with the $B=0$ case, we cannot say that the
Rashba electron has in the states (\ref{eq:r_spinor}) the spin projection
$\pm 1/2$ onto the direction $\alpha\vpi\times{\bf n}+\omega_{c}^{}\delta
{\bf n}$, because the components of the kinematic momentum operator $\vpi$
are not commuting motion integrals. Nevertheless, this interpretation
makes sense in the quasiclassical limit, when one can speak about the electron
path in a magnetic field. Namely, the quantum number $s=\pm 1$ determines
the value of the spin projection on the instant direction of $\alpha\vpi
\times{\bf n}+\omega_{c}^{}\delta{\bf n}$ that changes along the
quasiclassical electron path. Thus, the spin configurations of the Rashba
electron states form {\it vortices} in the $XY$--plane with center at the
origin.

The conductivity tensor $\hat{\sigma}$ of the considered system has just
one independent circularly polarized component $\sigma=\sigma_{xx}^{}+
i\sigma_{yx}^{}$. In the one--electron approximation, it has the form
\cite{gerhar_1975}
\begin{eqnarray}
\label{eq:kubo}
\sigma&\!\!=&\!\!\sigma_{}^{I}+\sigma_{}^{II}\nonumber\\
&\!\!=&\!\!\frac{e_{}^{2}}{8\pi}\Tr\,V_{+}^{}\!\Bigg[\!\left.\Big[
2\Phi_{EE}^{RA}-\Phi_{EE}^{RR}-\Phi_{EE}^{AA}\Big]\right|_{E=E_{F}^{}}^{}\!+
\nonumber\\
&\!\!&\!\!+\int_{-\infty}^{E_{F}^{}}\!\left.\big(\partial_{E}-\partial_{E_{}'}
\big)\!\Big[\Phi_{EE_{}'}^{AA}-\Phi_{EE_{}'}^{RR}\Big]\right|_{E_{}'=E}^{}\!
{\rm d}E\Bigg]\!.
\end{eqnarray}
Here, $\Phi_{EE_{}'}^{XY}=\big\langle\hat{G}_{}^{X}(E)V_{-}^{}\hat{G}_{}^{Y}
(E_{}')\big\rangle$ is the current vertex operator; $V_{\pm}^{}=V_{x}^{}\pm
iV_{y}^{}=v_{\pm}^{}\pm 2i\alpha\sigma_{\pm}^{}$ are circularly polarized
components of the full velocity operator [the corresponding components
$\vsigma$ are defined as $\sigma_{\pm}^{}=(\sigma_{x}^{}\pm i\sigma_{y}^{})/2$],
where the last term occurs due to SOI (\ref{eq:model_h}). $\hat{G}_{}^{R(A)}(E)
=1/(E-{\cal H}-U\pm i0)$ is the resolvent (retarded $(R)$ or advanced $(A)$) of
the Hamiltonian (\ref{eq:model_h}), and angular brackets $\langle\ldots\rangle$
denote the averaging over the random field $U$ configurations. Finally, the
symbol $\partial_{E}^{}$ denotes the derivative with respect to energy $E$.

\section{One--electron Green function}

By definition, the one--particle GF is the averaged resolvent of the
Hamiltonian (\ref{eq:model_h}) $\langle\hat{G}_{}^{R(A)}(E)\rangle=\langle
1/(E-{\cal H}-U\pm i0)\rangle$. It is connected with the electron self--energy
operator $\hat{\Sigma}_{}^{R(A)}(E)$ by the relation $(X=R,A)$
\begin{equation}
\label{eq:gf_def}
\langle\hat{G}_{}^{X}(E)\rangle=\left[\begin{array}{cc}
\langle G_{\uu}^{X}(E)\rangle & \langle G_{\ud}^{X}(E)\rangle\\
\langle G_{\du}^{X}(E)\rangle & \langle G_{\dd}^{X}(E)\rangle
\end{array}\right]=\frac{1}{E-{\cal H}-
\hat{\Sigma}_{}^{X}(E)}\,.
\end{equation}

The direct employment of the eigenspinors (\ref{eq:r_basis}) for calculation
of (\ref{eq:gf_def}), or kinetic and thermodynamic properties of the Rashba
system in a strong magnetic field leads to very complicated expressions. One
is forced almost from the first steps either to turn to numerical calculations
\cite{wang_etal_2003,averk_etal_2005,wang_etal_2005}, or to make simplifying
approximations like the momentum--independent spin--splitting energy
\cite{lange_etal_2004}. This makes more difficult the interpretation of the
results obtained in such a way, as well as the understanding of the whole
physical picture. But it turns out that the GF of the free $(U=0)$ Rashba
system is expressed {\it exactly} through the GF of the "ideal"\, electron in a
magnetic field. This opens up new possibilities for analytical studies of
the considered system. Indeed, it is easy to check that the Hamiltonian of the
free Rashba systems can be presented in the following form
\begin{equation}
\label{eq:h_rd_connect}
{\cal H}={\cal H}_{0}^{}+\nu\sqrt{2m\alpha_{}^{2}{\cal H}_{0}^{}
+\omega_{c}^{2}\delta_{}^{2}}\,.
\end{equation}
Here $\nu$ is the helicity operator defined in Eq.~(\ref{eq:helic_oper}),
${\cal H}_{0}^{}$ is the Hamiltonian of the "ideal"\, electron
(\ref{eq:ideal_h}).

The substitution of the Hamiltonian (\ref{eq:h_rd_connect}) into
the resolvent $\hat{G}(E)=(E-{\cal H})_{}^{-1}$ gives, after some
simple algebra, the following result (here and below, we drop
superscripts $R(A)$, if this does not lead to misunderstandings.
Sometimes, for brevity of notations, we shall not write explicitly
the energy arguments of the resolvents or GF's.)
\begin{equation}
\label{eq:step_one}
\hat{G}(E)=\frac{E-{\cal H}_{0}^{}+\big[\alpha(\vpi\times{\bf n})+
\omega_{c}^{}\delta{\bf n}\big]\cdot{\vsigma}}{(E+m\alpha_{}^{2}-
{\cal H}_{0}^{})_{}^{2}-\ds\frac{1}{4}\Omega_{B}^{2}}\,,
\end{equation}
where
\begin{equation}
\label{eq:spin_preces}
\Omega_{B}^{}=2\sqrt{2m\alpha_{}^{2}E+m_{}^{2}\alpha_{}^{4}+\omega_{c}^{2}
\delta_{}^{2}}=\sqrt{\Omega_{}^{2}+4\omega_{c}^{2}\delta_{}^{2}}\,.
\end{equation}
The quantity $\Omega_{B}^{}$ is equal to the magnetic field--dependent frequency
of the spin precession of the electron with energy $E$ that is responsible for
the Dyakonov --- Perel spin relaxation mechanism \cite{dyak_etal_1971a};
$\Omega$ is the same frequency in the absence of a magnetic field. It should be
noted that the same representation of the one--electron GF can be also obtained
for a system with the momentum--linear Dresselhaus SOI. For example, in the case
of a $[001]$--grown quantum well based on the
${\rm A}_{\rm III}^{}{\rm B}_{\rm V}^{}$ semiconductors, it is sufficient to
replace $\vpi\to\tilde{\vpi}=(\pi_{y}^{},\pi_{x}^{})$ in the definition of the
helicity operator (\ref{eq:helic_oper}), change the sign before the Zeeman term
($g_{0}^{}=-2$!) in the Hamiltonian of the "ideal"\,electron (\ref{eq:ideal_h}),
and, finally, to redefine the parameter $\delta\to\delta_{D}^{}=(g+2)/4$.

The denominator of the right--hand side of Eq.~(\ref{eq:step_one}) depends
on the "ideal"\, electron Hamiltonian alone. Expanding this expression
into the partial fractions, we obtain the desired representation of the
one--electron GF of the free Rashba system
\begin{eqnarray}
\label{eq:gf_repres}
&\!\!\!\hat{G}&\!\!\!\!(E)=\nonumber\\
&\!\!\!=&\!\!\!\frac{1}{2\Omega_{B}^{}}\sum_{s=\pm 1/2}\frac
{\Omega_{B}^{}+4s\big[m\alpha_{}^{2}-\omega_{c}^{}\delta_{}^{}\sigma_{z}^{}
-\alpha_{}^{}(\vpi\times{\bf n})\cdot\vsigma\big]}{E+m\alpha_{}^{2}
+s\Omega_{B}^{}-{\cal H}_{0}^{}}\nonumber\\
&\!\!\!=&\!\!\!\sum_{s=\pm 1/2}\left[\Phi_{s}^{}-2s\frac{\alpha_{}^{}(\vpi
\times{\bf n})\cdot\vsigma}{\Omega_{B}^{}}\right]\hat{G}(E+m\alpha_{}^{2}
+s\Omega_{B}^{})\,.\nonumber\\
\end{eqnarray}
We use here the same notation ($\hat{G}$) for the GF of the Rashba electron
and for the GF of the "ideal"\, electron. However, this does not lead to
confusion since the latter depends always on the energy arguments like
$E+m\alpha_{}^{2}+s\Omega_{B}^{}$ etc.

It is important that the same representation can be obtained for the averaged
resolvent of the Rashba system in the SCBA. We restrict ourselves here to an
approximation in which the electron self--energy operator is diagonal in the
spin space. Then, the SCBA equation for $\Sigma_{}^{X}(E)$ has the following
form
\begin{equation}
\label{eq:scba_def}
\hat{\Sigma}(E)=W\langle\Sp\hat{G}(E)\rangle=
\left[\begin{array}{cc}
\Sigma_{\uu}^{}(E) & 0\\
0 & \Sigma_{\dd}^{}(E)\end{array}\right].
\end{equation}
Here $\Sp$ denotes the trace only over the spatial degrees of freedom; $W=
n_{I}^{}U_{0}^{2}$, where $n_{I}^{}$ is the impurity concentration, $U_{0}^{}$
is the magnitude of the pointlike potential of an isolated impurity.
Therefore, it is sufficient
to make everywhere in Eq.~(\ref{eq:gf_repres}) the following substitutions
\begin{equation}
\label{eq:substitution}
E\,\to\,E-\Sigma_{e}^{}(E)\,\qquad g\omega_{c}^{}\,\to\,g\omega_{c}^{}+
4\Sigma_{o}^{}(E)
\end{equation}
to obtain the desired representations for the averaged GF's in the SCBA. Here
$\Sigma_{e(o)}^{}(E)=\big[\Sigma_{\uu}^{}(E)\pm\Sigma_{\dd}^{}(E)\big]/2$
are the even and odd parts of the electron self--energy. The first
($\Sigma_{e}^{}=\Delta_{e}^{}\pm i/2\tau_{e}^{}$) describes the perturbation
(shift $\Delta_{e}^{}$ and broadening $1/\tau_{e}^{}$) of the one--electron
energy levels by a random field. The real part of $\Sigma_{o}^{}=
\Delta_{o}^{}\pm i/2\tau_{o}^{}$ defines the renormalization of the Zeeman
coupling (\ref{eq:substitution}), while its imaginary part $\propto 1/
\tau_{o}^{}$ makes a contribution to the overall broadening of the one--electron
energy levels. As a result, we obtain a expression like Eq.~(\ref{eq:gf_repres})
for the averaged GF, where
\begin{equation}
\label{eq:ideal_gf}
\hat{G}_{}^{R(A)}(E+m\alpha_{}^{2}+s\Omega_{B}^{})=\frac{1}{E+m\alpha_{}^{2}
+s\Omega_{B}^{}-{\cal H}_{0}^{}\pm\ds\frac{i}{2\tau_{s}^{}}}
\end{equation}
is the averaged retarded (advanced) GF of the "ideal"\, electron, and
\begin{eqnarray}
\label{eq:renorm}
\Omega_{B}^{}&\!\!\!=&\!\!\!\frac{1}{2}\big(\Omega_{B}^{R}+\Omega_{B}^{A}\big)
\,,\nonumber\\
\frac{1}{\tau_{s}^{}}&\!\!\!=&\!\!\!\frac{1}{{\tau}_{e}^{}}-is\big(
\Omega_{B}^{R}-\Omega_{B}^{A}\big)=\left(1+s\frac{4m\alpha_{}^{2}}
{\Omega_{B}^{}}\right)\frac{1}{\tau_{e}^{}}+s\frac{4\omega_{c}^{}\delta}
{\Omega_{B}^{}}\frac{1}{\tau_{o}^{}}\nonumber\\
\end{eqnarray}
are the disorder--modified frequency of the spin precession
(\ref{eq:spin_preces}) and the inverse life time of an electron in the $s$--th
spin--split subband. As usual, we do not take explicitly into consideration in
(\ref{eq:ideal_gf}) the one--electron energy levels shift $\Delta_{e}^{}$ that
is absorbed by the normalization condition, but we mean here that the odd shift
$\Delta_{o}^{}$ is included in the definition of the effective $g$--factor in
accordance with (\ref{eq:substitution}). The explicit allowance for the Zeeman
coupling renormalization is particularly important in the SdH oscillations
regime.

\section{Density of states and self--energy}

We first consider the calculation of the DOS $n(E)=\Im\langle\Tr\hat
{G}_{}^{A}(E)\rangle/\pi$ using the above--obtained expression for the
one--particle GF (\ref{eq:gf_repres}). Here, the symbol $\Tr$ denotes the trace
over the spatial and spin degrees of freedom. For the sake of simplicity, we
shall deal with the case of large filling numbers $(E\gg\omega_{c}^{})$.
Calculating the trace of resolvent (\ref{eq:gf_repres}) over the spatial and
spin degrees of freedom, we obtain the following expression for the DOS
\begin{eqnarray}
\label{eq:dos_scba}
n(E)&\!\!\!=&\!\!\!\sum_{s=\pm 1/2}\frac{m_{s}^{}}{m}n_{}^{(0)}\big[E+
m\alpha_{}^{2}+s(\Omega_{B}^{}\pm\omega_{c}^{})\big]\nonumber\\
&\!\!\!=&\!\!\!\sum_{s=\pm 1/2}\frac{m_{s}^{}}{m}n_{s}^{(0)}(E)\,.
\end{eqnarray}
Here, we take into account that the DOS of a spinless electron in an orthogonal
magnetic field $n_{}^{(0)}(E)$ satisfies $n_{}^{(0)}(E)=n_{}^{(0)}
(E\pm\omega_{c}^{})$ at large filling factors $(E\gg\omega_{c}^{})$. The sign
before $\omega_{c}$ is chosen in such a way as to ensure the right--hand limit
$s(\Omega_{B}^{}\pm\omega_{c}^{})\to\pm sg\omega_{c}^{}/2$, as the spin--orbit
coupling approaches zero. The effective mass $m_{s}^{}$ in the $s$--th subband
is defined as
\begin{equation}
\label{eq:eff_mass}
m_{s}^{}=m\left(1+s\frac{4m\alpha_{}^{2}}{\Omega_{B}^{}}\right)=
m\partial_{E}^{}(E+s\Omega_{B}^{})\,.
\end{equation}
In the considered case this expression coincides with the usual definition of
the transport and cyclotron effective masses in the isotropic nonparabolic band
\cite{tsidil_1978}.

In full accordance with the two--subband model, the DOS
in Eq.~(\ref{eq:dos_scba}) is presented as a sum of partial contributions. Using
this expression for the DOS, we can obtain the analytical form of the equation
for the electron concentration $n=\int_{}^{E_{F}^{}}n(E){\rm d}E$ that is the
normalization condition for the Fermi level determination.
So, a more correct expression for the DOS is needed, which adequately describes
its behavior not only in the vicinity of the Fermi level $E_{F}^{}(>0)$, but also
near the lower boundary of the spectrum. For example, at $B=0$ we have
\begin{equation}
\label{eq:dos_zero}
n(E)=\frac{m}{\pi}\left\{\begin{array}{cl}
\ds\frac{m\alpha}{\sqrt{2mE+m_{}^{2}\alpha_{}^{2}}}\,, & -\ds\frac{1}{2}
m\alpha_{}^{2}<E<0\,,\\
1\,, & E\geq 0\,.\end{array}\right.
\end{equation}
In the energy interval $-m\alpha_{}^{2}/2<E<0$, the DOS is formed by the states
of the lower spin-split subband and has the typical one--dimensional behavior. 
Integrating (\ref{eq:dos_zero}) between $-m\alpha_{}^{2}/2$ and
$E_{F}^{}$, we obtain
\begin{equation}
\label{eq:zero_norm}
n=\frac{m}{\pi}\big(E_{F}^{}+m\alpha_{}^{2}\big)=\frac{m}{\pi}E_{0}^{}\,,
\quad E_{F}^{}>0\,.
\end{equation}
Thus, the energy $E_{0}^{}=E_{F}^{}+m\alpha_{}^{2}$ corresponds to the Fermi
level in the absence of SOI. Notice that the partial electron concentrations
$n_{s}^{}=m(E_{0}^{}+s\Omega_{B}^{})/2\pi$ depend nonlinearly on the Fermi
energy, in contrast to $n$ (\ref{eq:zero_norm}). The correction to the Fermi
energy in (\ref{eq:zero_norm}) comes from the low-energy tail of the DOS
(\ref{eq:dos_zero}). Of course, the difference between $E_{0}^{}$ and $E_{F}^{}$
is small for weak SOI ($m\alpha_{}^{2}\ll E_{F}^{}$).
However, as shown below, it is of crucial importance to take it into account for
a correct interpretation of the spin--orbit interaction effect on both the
conductivity in the absence of a magnetic field, and the SdH oscillations.

The representation (\ref{eq:dos_scba}) allows one to obtain a simple analytical
expression for the DOS that holds good up to the quantizing fields region
($\omega_{c}^{}\tau\gtrsim 1$). Indeed, the DOS of a spinless electron in the
large filling factors region ($E\gg\omega_{c}^{}$) has the form
\begin{equation}
\label{eq:dos_spinless}
n_{}^{(0)}(E)=\frac{m}{2\pi}\frac{\sinh\ds\frac{\pi}{\omega_{c}^{}\tau}}
{\cosh\ds\frac{\pi}{\omega_{c}^{}\tau}+\cos 2\pi\ds\frac{E}{\omega_{c}^{}}}
\end{equation}
Inserting Eq.~(\ref{eq:dos_spinless}) into Eq.~(\ref{eq:dos_scba}), we
obtain for the oscillating part of the DOS the following expression
\begin{eqnarray}
\label{eq:dos_sdh}
\Delta n(E_{F}^{})=\frac{2m}{\pi}\exp\left(-\frac{\pi}{\omega_{c}^{}\tau}
\right)\left[\cos 2\pi\frac{E_{0}^{}}{\omega_{c}^{}}\cos\pi\frac{\Omega_{B}^{}}
{\omega_{c}^{}}-\right.\nonumber\\
-\left.\frac{2m\alpha_{}^{2}}{\Omega_{B}^{}}\sin 2\pi\frac{E_{0}^{}}
{\omega_{c}^{}}\sin\pi\frac{\Omega_{B}^{}}{\omega_{c}}\right]
\end{eqnarray}
that is valid in the magnetic fields region under consideration. The second term
in Eq.~(\ref{eq:dos_sdh}) appears due to the difference between the effective
masses $m_{s}^{}$ (\ref{eq:eff_mass}).

It follows from (\ref{eq:dos_sdh}) that energy $E_{0}^{}$  defines the main
period of the SdH oscillations. In other words, in the large filling factors region
($E_{F}^{}\gg\omega_{c}^{}$) the SdH period is determined by the total
electron concentration (see Eq.~(\ref{eq:zero_norm})), redardless of the
spin--orbit interaction magnitude. On the other hand, the period of the SdH
oscillation beatings (\ref{eq:dos_sdh}) (see Fig.~1(a)) defines the spin precession
frequency (\ref{eq:spin_preces}) which depends on the magnetic field even in
the absence of Zeeman splitting ($g=0$).

In the case of weak SOI ($\Omega\ll E$), the oscillations of the DOS are
determined completely by the first term in Eq.~(\ref{eq:dos_sdh}). Then, the
location of the $k$-th node of beatings is determined by the condition
\begin{equation}
\label{eq:node}
B_{k}^{}=\frac{2mc}{|e|}\frac{\Omega}{\sqrt{(2k+1)_{}^{2}-(g-2)_{}^{2}}}\,.
\end{equation}

This limit was considered in Ref.~\onlinecite{taras_etal_2002}. Unlike
the results of that work, the above--obtained equations still
stand in the case of strong SOI, where it is important to take
account of the difference $E_{0}^{}$ and $E_{F}^{}$ for correctly
defining the SdG oscillation period. In addition, we have taken
into account the Zeeman splitting of the electron spectrum that
allows to describe more correctly the oscillation pattern. For
example, the Eq.~(\ref{eq:node}) allows us to determine both the
spin--orbit, $\alpha$, and Zeeman, $g$, couplings by the measured
locations of two different nodes (see the upper curve in
Fig.~1(a)). On the other hand, the spin precession frequency
$\Omega_{B}^{}$ approaches $|\delta|\omega_{c}^{}$ as the magnetic
field $B$ increases. Therefore, in this case a gradual transition
from the beatings of the SdH oscillations to the familiar Zeeman
splitting of the oscillating peaks should be observed. The
beginning of this transition can be seen on the lower curve in
Fig.~1(a).

\begin{figure}[t!]
\begin{center}
\includegraphics[scale=0.56]{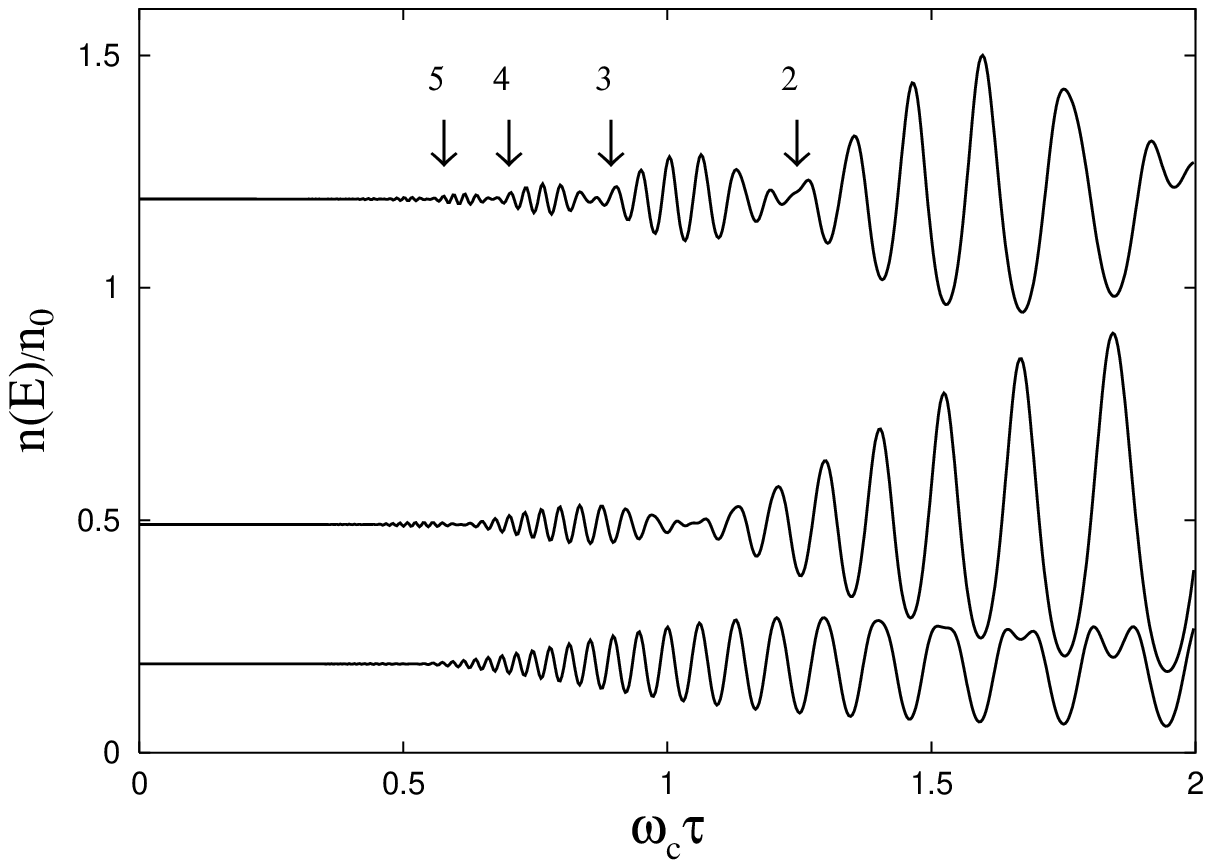}

\includegraphics[scale=0.56]{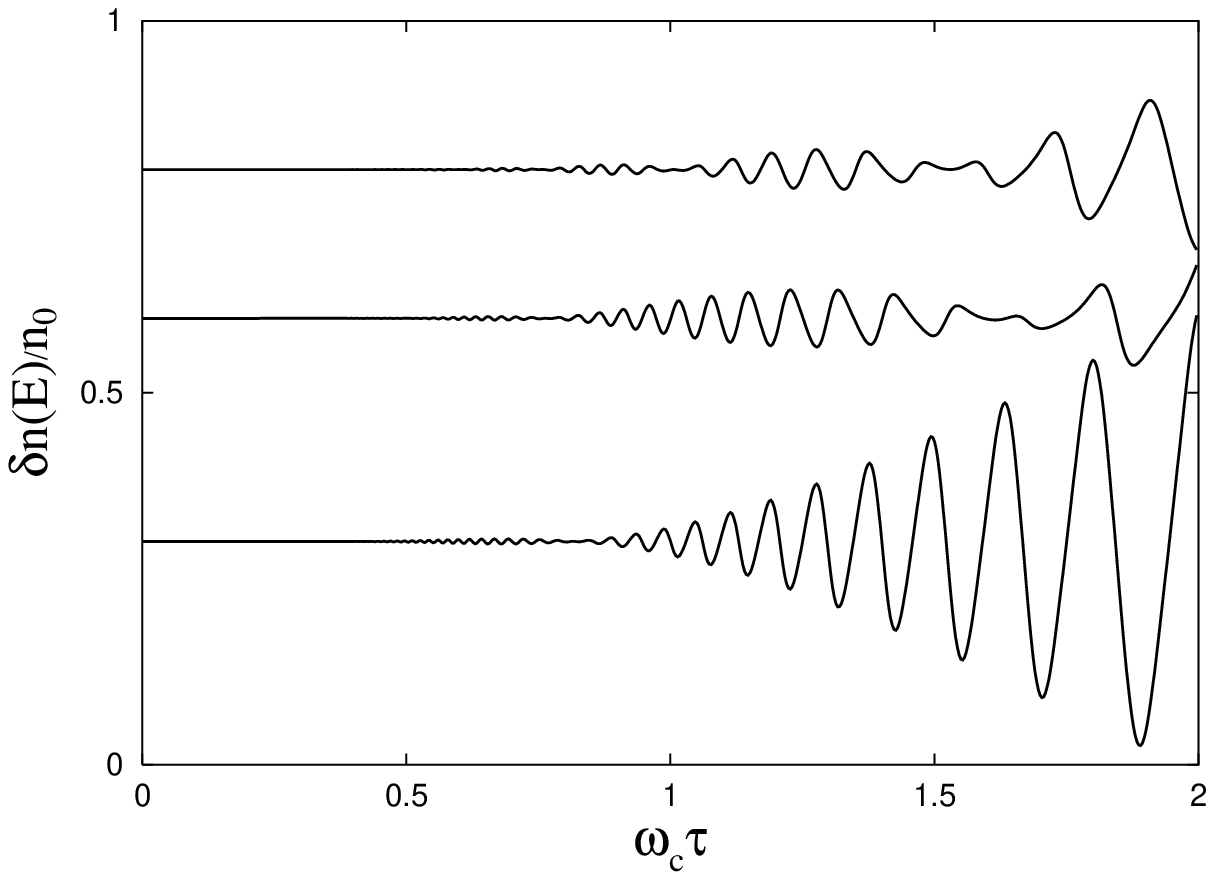}
\caption{Plots of the SdH oscillations of the total DOS (upper panel) and
of the difference of the partial DOS's (bottom panel) of the 2D Rashba
system at fixed $g=2.8$, and $k_{F}^{}l=35.0$, and different $\Omega\tau=
3.0;\;1.5;\;0.75$ (up to down). The arrows point the nodes location with
their numbers $k=2,3,4,5$ that are calculated with Eq.~(\ref{eq:node}).}
\end{center}
\end{figure}
\begin{figure}[t!]
\begin{center}
\includegraphics[scale=0.56]{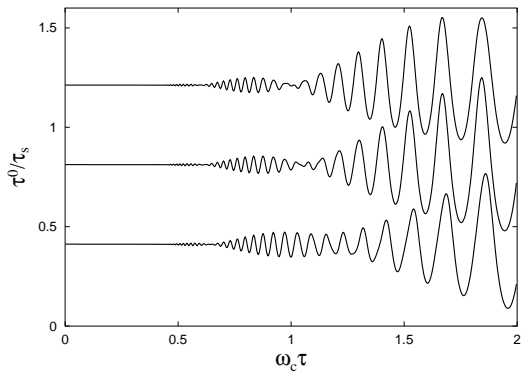}
\caption{Plots of the SdH oscillations of the inverse life time of
one--electron states in the $s$--th spin--splitted subband at different
values of Zeeman factor $g=1.8;\;1.0;\;0.2$ (up to down), and fixed values
of $k_{F}^{}l=35.0$, and $\Omega\tau=1.5$.}
\end{center}
\end{figure}

Another important characteristic of the one--electron states of the 2D Rashba
system is the difference of the partial DOS's with opposite spin projections
onto the $OZ$--axis
\begin{equation}
\label{eq:ddos}
\delta n(E)=n_{\uu}^{}(E)-n_{\dd}^{}(E)=-\frac{4\omega_{c}^{}\delta}
{\Omega_{B}^{}}\sum_{s=\pm 1/2}sn_{s}^{(0)}(E)
\end{equation}
This quantity is proportional to the derivative of the transverse
spin magnetization with respect to energy $E$ and, therefore, it
enters in the definition of the effective concentrations of
current carriers in the dissipative part of the 2D Rashba system
conductivity in an orthogonal magnetic field (see the next
section).

Evidently, the difference of the partial DOS's (\ref{eq:ddos}) vanishes in the
region of classical magnetic fields ($\omega_{c}^{}\tau\ll 1$), but it plays an
important role in the SdH oscillations regime. In the case of large filling
factors, the oscillating behavior of this quantity is described by the following
expression
\begin{equation}
\label{eq:ddos_sdh}
\delta n(E_{F}^{})=\frac{2m}{\pi}\frac{2\omega_{c}^{}\delta}{\Omega_{B}^{}}
\exp\left(-\frac{\pi}{\omega_{c}^{}\tau}\right)\sin 2\pi\frac{E_{0}^{}}
{\omega_{c}^{}}\sin\pi\frac{\Omega_{B}^{}}{\omega_{c}^{}}\,.
\end{equation}
Unlike the total DOS (\ref{eq:dos_sdh}), this expression contains just one
oscillating term, because $\delta n(E)$ does not depend on the effective masses
$m_{s}^{}$ (\ref{eq:eff_mass}). Indeed, the difference of the partial
DOS's $\delta n(E)$ is non zero, which is  entirely due to the spin degrees of
freedom of the electrons. The typical SdH oscillation patterns of $\delta n(E)$
are depicted in Fig.~1(b).

Now, let us turn to the discussion of the electron life time $\tau_{s}^{}$
in the $s$--th spin--split subband which is defined, according to
Eq.~(\ref{eq:renorm}), by the imaginary parts of the even and odd self--energies
$\Sigma_{e(o)}^{}$. In other words, the total life time of the one--electron
states $\tau_{s}^{}$ is determined by the sum of the weighted relaxation
rates of the orbital and spin degrees of freedom. The first term in this
expression is proportional to the above--considered total DOS, hence its
magnetic--field dependence coincides up to the scale factor with the patterns
shown in Fig.~1(a). Of particular interest is the last term in
Eq.~(\ref{eq:renorm}) stemming from the Zeeman coupling renormalization. It is
proportional to the difference of the partial DOS's (\ref{eq:ddos}) and,
therefore, plays an important role in the SdH oscillation regime, as shown in
Fig.~2. Notice that the beatings of the SdH oscillations are supressed with the
increase of the relative magnitude of the second term in Eq.~(\ref{eq:renorm}).
Indeed, Eq.~(\ref{eq:node}) determines the location of the beating loops of the
oscillation instead of the nodes. Thus, the broadening of the Zeeman levels
leads to observable supression of the beatings of the electron life time
$\tau_{s}^{}$ oscillations.

\section{Conductivity}

The general expression for the conductivity (\ref{eq:kubo})
consists of two different terms. The first of them describes the
contribution of the electrons at the Fermi level, the second one
contains the contributions of all filled states below the Fermi
level. We begin the calculation of the conductivity with the last
term of (\ref{eq:kubo}) $\sigma_{}^{II}$. First of all, it is pure
imaginary and, therefore, makes a contribution in the Hall
conductivity alone. St\u{r}eda \cite{streda_etal} was first to
show that, for spinless electrons, this part of the conductivity
is equal to
\begin{equation}
\label{eq:streda}
\sigma_{}^{II}=i|e|c\left(\frac{\partial n}{\partial B}\right)_{E_{F}^{}}^{}\,,
\end{equation}
where $n$ is the electron concentration. It should be pointed out that
Eq.~(\ref{eq:streda}) is {\it exact}, and with the thermodynamic Maxwell
relation $\sigma_{}^{II}$ can be expressed through
$(\partial M/\partial E)_{B}^{}$, where $M$ is the orbital magnetization of the
electron gas. Detailed discussion of $\sigma_{}^{II}$ and its physical
interpretation can be found in survey \cite{pruisken}.

This result is extended immediately to the electron systems with SOI.
Following St\u{r}eda's argument, it can be shown that the part
$\sigma_{}^{II}$ of the 2D Rashba system conductivity is expressed as
\begin{eqnarray}
\label{eq:soi_streda}
\sigma_{}^{II}=i|e|c\left[\left(\frac{\partial n}{\partial B}
\right)_{E_{F}^{}}^{}-\left(\frac{\partial M_{p}^{}}{\partial E}\right)_{B}^{}
\right]\,,
\end{eqnarray}
where $M_{p}^{}$ is the spin magnetization of the electron gas.
It follows that in the general case $\sigma_{}^{II}$ is determined only by the
diamagnetic part of the electron gas magnetization. By direct differentiation of
the electron concentration $n$ with respect to the magnetic field induction $B$
we obtain
\begin{widetext}
\begin{equation}
\label{eq:dn_db}
\left(\frac{\partial n}{\partial B}\right)_{E_{F}^{}}^{}=\frac
{n}{B}+\int^{E_{F}^{}}\Tr\frac{\partial}{\partial B}\big(G_{}^{A}-G_{}^{R}\big)
\frac{{\rm d}E}{2\pi i}
=\frac{n}{B}-\int^{E_{F}^{}}\Tr\left[\frac
{\partial{\cal H}}{\partial B}\frac{\partial\hat{G}_{}^{A}}{\partial E}+
\frac{\partial\Sigma_{}^{A}}{\partial B}\frac{\partial\hat{G}_{}^{A}}
{\partial E}-\frac{\partial\Sigma_{}^{A}}{\partial E}\frac
{\partial\hat{G}_{}^{A}}{\partial B}-h.c.\right]\frac{{\rm d}E}{2\pi i}\,,
\end{equation}
\end{widetext}
where the symbol $h.c.$ denotes the Hermitian conjugate the terms. Considering
that in the SCBA $\Sigma=W\Tr\hat{G}/2$, it is easily seen that the
expression in the square brackets is the total derivative with respect to energy.
As a result, integration by $E$ in (\ref{eq:dn_db}) is performed  explicitly, and
after some simple algebra the expression for $\sigma_{}^{II}$ takes on the form
\begin{equation}
\label{eq:sigma_streda}
\sigma_{}^{II}=i\frac{|e|c}{B}\big(n-n_{\perp}^{}\big)\,,
\end{equation}
where
\begin{equation}
\label{eq:n_p}
n_{\perp}^{}=\frac{1}{2\pi i}\Tr\left(E_{0}^{}-\frac{1}{4}g\omega_{c}^{}
\sigma_{z}^{}\right)\big(\hat{G}_{}^{A}-\hat{G}_{}^{R}\big)\,.
\end{equation}

Eq.~(\ref{eq:sigma_streda}) is a generalization of the known SCBA expression for
$\sigma_{}^{II}$ in the case of spinless electrons \cite{pruisken}. Particularly,
the quantity $n_{\perp}^{}$ (\ref{eq:n_p})
is the counterpart of the familiar parameter $n_{\perp}^{}=En_{}^{(0)}(E)$ that
stands for the current carrier concentration in the dissipative part of the
conductivity tensor of spinless 2D--electrons in a magnetic field in the SCBA
\cite{gerhar_1975}. In the classical (i.e. nonquantizing, $\omega_{c}\tau\ll 1$)
magnetic fields region, it is equal to the total electron concentration $n$.

\begin{figure}[h!]
\includegraphics[scale=0.634]{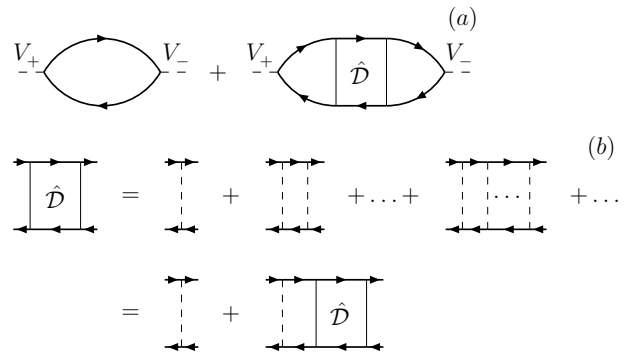}
\caption{(a) The diagrams depicting the conductivity in the ladder
approximation; (b) The impurity ladder series in the
particle--hole channel (diffuson).}
\end{figure}

Now, we turn to the first term in the conductivity (\ref{eq:kubo}). It is quite
easy to show, by identical transformations, that
\begin{equation}
\label{eq:rr_aa_part}
\frac{e_{}^{2}}{8\pi}\Tr V_{+}^{}\big[\Phi_{EE}^{AA}+\Phi_{EE}^{RR}\big]=-\frac
{e_{}^{2}}{4\pi m}\Tr\big(\hat{G}_{}^{A}+\hat{G}_{}^{R}\big)\,.
\end{equation}
The main contribution to the dissipative part of the conductivity is
proportional to the current vertex $\Phi_{EE}^{RA}$ in Eq.~(\ref{eq:kubo}).
If we accept the SCBA (\ref{eq:scba_def}) for the electron self--energy
$\hat\Sigma$, we must evaluate this part of conductivity in the ladder
approximation in order to satisfy the particle conservation law (see. Fig.~3).

Replacing, as a first approximation, $\Tr V_{+}^{}\Phi_{EE}^{RA}\to\Tr V_{+}^{}
\hat{G}_{}^{R}V_{-}^{}\hat{G}_{}^{A}$, we obtain the "bare"\, conductivity
\begin{equation}
\label{eq:s_bare}
\sigma_{\rm bare}^{I}=\frac{e_{}^{2}}{4\pi}\Tr\left[V_{+}^{}\hat{G}_{}^{R}
V_{-}^{}\hat{G}_{}^{A}+\frac{1}{m}\big(\hat{G}_{}^{A}+\hat{G}_{}^{R}\big)\right]
\,,
\end{equation}
depicted by the first diagram in Fig.~3(a). In what follows, for simplicity, we neglect
everywhere the odd part of the electron self--energy $\Sigma_{o}^{}$ (\ref{eq:renorm}).
In this approximation one can obtain the following equations
\begin{widetext}
\begin{equation}
\label{eq:identity}
\begin{split}
\left(\omega_{c}^{}+\frac{i}{\tau_{e}^{}}\right)\hat{G}_{}^{R}V_{-}^{}
\hat{G}_{}^{A}&=V_{-}^{}\hat{G}_{}^{A}-\hat{G}_{}^{R}V_{-}^{}+2i\alpha\left[
\sigma_{-}^{}\hat{G}_{}^{A}-\hat{G}_{}^{R}\sigma_{-}^{}-\frac{i}{\tau_{e}^{}}
\hat{G}_{}^{R}\sigma_{-}^{}\hat{G}_{}^{A}\right]\,,\\
\left(\omega_{c}^{}+\frac{i}{\tau_{e}^{}}\right)\hat{G}_{}^{A}V_{+}^{}
\hat{G}_{}^{R}&=\hat{G}_{}^{A}V_{+}^{}-V_{+}^{}\hat{G}_{}^{R}-2i\alpha\left[
\hat{G}_{}^{A}\sigma_{+}^{}-\sigma_{+}^{}\hat{G}_{}^{R}-\frac{i}{\tau_{e}^{}}
\hat{G}_{}^{A}\sigma_{+}^{}\hat{G}_{}^{R}\right]\,.
\end{split}
\end{equation}
\end{widetext}
Using these relations, we perform a series of transformations of the "bare"\,
conductivity (\ref{eq:s_bare}), neglecting the terms being small in parameter
$1/E_{F}^{}\tau$ and lying beyond the accuracy of the ladder approximation.
Omitting intermediate transformations of purely technical character, the final
result can be written as
\begin{equation}
\label{eq:bare}
\sigma_{\rm bare}^{I}=\frac{e_{}^{2}\tau_{e}^{}}{m}\frac{1}
{1-i\omega_{c}^{}\tau_{e}^{}}\left[n_{\perp}^{}-\frac{2m\alpha_{}^{2}n_{F}^{}}
{1-i\omega_{c}^{}\tau_{e}^{}}(1-P)\right]\,,
\end{equation}
where $n_{F}^{}=\Tr(\hat{G}_{}^{A}-\hat{G}_{}^{R})/4\pi i$ is the DOS at the
Fermi level per spin;
\begin{equation}
\label{eq:p}
P=W\Tr\sigma_{+}^{}\hat{G}_{}^{R}\sigma_{-}^{}\hat{G}_{}^{A}\,.
\end{equation}

The ladder correction to the conductivity shown by the second diagram in Fig.~3(a)
is presented by the following analytical expression
\begin{equation}
\label{eq:s_ladder}
\Delta\sigma_{\rm lad}^{I}=\frac{e_{}^{2}}{4\pi}\tr\left[\Sp\big(\hat{G}_{}^{A}
V_{+}^{}\hat{G}_{}^{R}\big)\cdot\hat{\cal D}\cdot\Sp\big(\hat{G}_{}^{R}V_{-}^{}
\hat{G}_{}^{A}\big)\right]\,.
\end{equation}
The "bare"\, current vertices
$\Sp\big(\hat{G}_{}^{A(R)}V_{\pm}^{}\hat{G}_{}^{R(A)}\big)$ involved in
(\ref{eq:s_ladder}) can be expressed by the parameter $P$ (\ref{eq:p}) and
the relaxation time $\tau_{e}^{}$, using relations (\ref{eq:identity}). In the
main order in the small parameter $1/E_{F}^{}\tau$ they have the form
\begin{equation}
\label{eq:gvg}
\Sp\big(\hat{G}_{}^{A(R)}V_{\pm}^{}\hat{G}_{}^{R(A)}\big)=\mp i\sigma_{\pm}^{}
\frac{4\pi\alpha n_{F}^{}\tau_{e}^{}}{1-i\omega_{c}^{}\tau_{e}^{}}(1-P)\,.
\end{equation}

The diffuson $\hat{\cal D}$ depicted in Fig.~3(b) by a ladder series in the
particle--hole channel is a $4\times 4$--matrix in the representation of the
total spin of the electron--hole pair. The Bete--Salpeter equation for
$\hat{\cal D}$ (see the second row in Fig.~3(b)) has the same structure. 
However, as seen from (\ref{eq:s_ladder}) and (\ref{eq:gvg}), the
contribution to $\Delta\sigma_{\rm lad}^{I}$ comes from the scalar quantity
${\cal D}=\tr\sigma_{+}^{}\hat{\cal D}\sigma_{-}^{}$ being the projection of
the diffuson on the triplet state $|1,-1\rangle$ of the electron--hole pair.
Projecting the Bete--Salpeter equation on this state, we obtain a closed scalar
equation for ${\cal D}$ with the solution of the form
\begin{equation}
\label{eq:diffuson}
{\cal D}=\frac{W}{1-P}\,,
\end{equation}
where $P$ is defined in (\ref{eq:p}).

Substitution of (\ref{eq:diffuson}) and (\ref{eq:gvg}) into (\ref{eq:s_ladder})
yields the expression for the ladder correction to the conductivity which
exactly cancels the second term in the square brackets in (\ref{eq:bare}).
As a result, the final expression for the conductivity of the Rashba system
in a transverse magnetic field in the ladder approximation takes on the form
\begin{equation}
\label{eq:total_sigma}
\sigma=i\frac{|e|c}{B}\left[n-\frac{n_{\perp}^{}}{1-i\omega_{c}^{}\tau_{e}^{}}
\right]\,.
\end{equation}

Eq.~(\ref{eq:total_sigma}) looks as if the current were generated by charge
carriers of one type with mobility $\mu=|e|\tau_{e}^{}/m$ and concentration $n$.
This would be expected, because the conductivity tensor in the absence of
a magnetic field is diagonal in the original spin space ($\sigma_{\ud}^{}=
\sigma_{\du}^{}\equiv 0$) by virtue of the momentum parity of the GF's, and the
full conductivity is equal to $\sigma=\sigma_{\uu}^{}+\sigma_{\dd}^{}$
\cite{inoue_etal}. In the classical (nonquantizing) magnetic fields this
property is retained, since the difference of the partial DOS's with opposite
spin projections onto the $OZ$--axis (\ref{eq:ddos}) is equal to zero in this
region (see Fig.~1).

Of course, the simple structure of Eq.~(\ref{eq:total_sigma}) for the
conductivity breaks down in the region of sufficiently strong magnetic fields,
$\omega_{c}^{}\tau\gg 1$, where the contribution $1/\tau_{o}^{}$ to the inverse
life time of the one--electron states (\ref{eq:renorm}) cannot already be
neglected. As shown in the previous section, this may lead to in flattening of
the SdH oscillation beatings (see Fig. 2). Therefore we believe the
approximation used in this section $1/\tau_{o}^{}=0$ to be correct in the
magnetic field region where well--defined beating nodes of magnetooscillation
are observed.

\section{Results and discussion}

First of all, let us summarize briefly the main results obtained in this work.
We have shown that the eigenstates of the 2D Rashba electron in an orthogonal
magnetic field are characterized by a special motion integral
(\ref{eq:helic_oper}) that generalizes the notion of {\it helicity}
\cite{edelst_1990}. Using this fact, we have found the relation
(\ref{eq:gf_repres}) between the GF's of the 2D Rashba electron and the "ideal"\,
one that holds good for arbitrary orthogonal magnetic fields as well as for the
strong spin--orbit coupling. With the help of this relation, we have obtained,
in contrast to Refs.~\onlinecite{wang_etal_2003,lange_etal_2004,wang_etal_2005}, the
analytical expressions for the DOS in SCBA (\ref{eq:dos_scba}) and
for the magnetoconductivity in the ladder approximation (\ref{eq:total_sigma})
of the 2D Rashba system that are valid in a wide range from the classical
magnetic fields up to the quantizing ones ($\omega_{c}^{}\tau \gtrsim 1$). The
spin--orbit as well as the Zeeman splitting of the electron energy
are properly allowed for in these expressions, unlike the results
of Refs.~\onlinecite{taras_etal_2002,averk_etal_2005}. In particular, we
have obtained a simple expression (\ref{eq:node}) for the nodes
locations of the SdH oscillations beatings. We have shown that the
competition of the relaxation rates of the orbital and spin
degrees of freedom in the total inverse life time $1/\tau_{s}^{}$
of the one--electron states in the $s$--th subband leads to the
partial supression of beatings of the $1/\tau_{s}^{}$ SdH
oscillations.

We start the discussion of the results with the conductivity in the classical
magnetic fields region ($\omega_{c}^{}\tau_{e}^{}\ll 1$). In this case, it
follows immediately from Eqs. (\ref{eq:n_p}), and (\ref{eq:total_sigma}) that
the conductivity of a 2D Rashba system takes the usual Drude--Boltzmann form
\begin{equation}
\label{eq:drude_cond}
\sigma=\frac{\sigma_{D}^{}}{1-i\omega_{c}^{}\tau_{e}^{}}\,,
\end{equation}
where
\begin{equation}
\label{eq:zero_cond}
\sigma_{D}^{}=\frac{e_{}^{2}\tau_{e}^{}}{\pi}(E_{F}^{}+m\alpha_{}^{2})=
\frac{e_{}^{2}n\tau_{e}^{}}{m}\,,
\end{equation}
is the Drudian conductivity in absence of the magnetic field; $\tau_{e}^{}=1/mW$
is the life time of an one--electron state at $B=0$. It immediately follows that
in the ladder approximation the classical magnetoresistance of a 2D Rashba
system is zero, $\rho(B)=\rho_{D}^{}=1/\sigma_{D}^{}$, and the Hall coefficient
$R_{H}^{}=-1/|e|nc$.

Thus, in the ladder approximation the Rashba spin--orbit interaction has {\it no
effect at all} on the conductivity magnitude everywhere over the region of
classical magnetic fields, $\omega_{c}^{}\tau_{e}^{}\ll 1$ (including the case
of $B=0$). Note that the first relation for the Drudian conductivity
(\ref{eq:zero_cond}) formally coincides with that obtained in
Ref.~\onlinecite{inoue_etal}. However, as mentioned above, the
correction $m\alpha_{}^{2}$ to the Fermi energy does not lead to
an observable change in $\sigma$, since it is absorbed by the
normalization condition (\ref{eq:zero_norm}).

Let us proceed now to the discussion of the magnetotransport in the 2D Rashba
system in the the large filling factors $(E\gg\omega_{c}^{})$ region, where the
SCBA and the ladder approximation are applicable to the description of the
one--electron states and kinetic phenomena, respectively. As usual, we extract
in the linear approximation the oscillating parts of the conductivity that enter
through DOS into the effective concentration $n_{\perp}^{}$ (\ref{eq:n_p}) and
mobility $\mu$. As in the ladder approximation the conductivity
(\ref{eq:total_sigma}) has the form characteristic of a conductor with one type
of charge carriers, we can immediately use the expression obtained in
Ref.~\onlinecite{isih_etal_1986} for the oscillating parts of the longitudinal
resistance $\rho$ and the Hall coefficient $R_{H}^{}$
\begin{subequations}
\label{eq:final_sdh}
\begin{eqnarray}
\label{eq:rho_sdh}
\frac{\Delta\rho(B)}{\rho_{0}^{}}&\!\!=&\!\!2\frac{\Delta n(E_{F}^{})}
{n_{F}^{(0)}}\,,\\
\label{eq:rhoh_sdh}
\frac{\Delta R_{H}^{}(B)}{R_{H}^{0}}&\!\!=&\!\!\frac{1}
{\omega_{c}^{2}\tau_{e}^{2}}\frac{\Delta n(E_{F}^{})}{n_{F}^{(0)}}\,,
\end{eqnarray}
\end{subequations}
Here, $\rho_{0}^{}=1/\sigma_{D}^{}$ and $R_{H}^{0}=-1/|e|nc$ are the
resistance (see Eq.~(\ref{eq:zero_cond})) and Hall coefficient in zero magnetic
field, respectively. Thus, in the linear approximation the SdH oscillations
$\rho$ and $R_{H}^{}$ are entirely determined by the magnetic--field dependence of
the DOS, $\Delta n(E_{F}^{})$, the first harmonics of which are of the form
(\ref{eq:dos_sdh}). From this we can draw two conclusions that are of great importance
for an adequate interpretation of the SdH oscillation pattern in systems with
spin--orbit interaction.

First, the period of the SdH oscillations of $\rho$ and $R_{H}^{}$ is defined by energy
$E_{0}^{}=E_{F}^{}+m\alpha_{}^{2}$, and not by the Fermi energy $E_{F}^{}$, as was stated
in Refs.~\onlinecite{taras_etal_2002,averk_etal_2005}. From this and the normalization 
condition (\ref{eq:zero_norm}) it follows that the SdH oscillation period is related to 
the total charge carriers concentration $n$ by the well known formula
\begin{equation}
\label{eq:sdh_period}
\Delta\left(\frac{1}{B}\right)=\frac{|e|}{\pi cn}\,,
\end{equation}
that holds true irrespective of the magnitude of the spin--orbit interaction constant.

Second, the SdH oscillation beating period is defined by the magnetic--field--dependent
spin precession frequency $\Omega_{B}^{}=\sqrt{\Omega_{}^{2}+(g-2)_{}^{2}\omega_{c}^{2}}$
(see. Eq.~(\ref{eq:spin_preces})). From this it follows a simple equation for the beating
nodes location $B_{k}^{}$ (\ref{eq:node}) that allows the constants $\alpha$ and $g$ to be
found from the measured $B_{k}^{}$ values. This is illustrated Fig.~4 which presents the
results of least--squares fitting of Eq.~(\ref{eq:node}) in variables $B_{}^{-2}$,
$(2k+1)_{}^{2}$ to the measured \cite{das_etal} locations of the SdH oscillation beating
nodes of the longitudinal resistance of a ${\rm In}_{x}^{}{\rm Ga}_{1-x}^{}
{\rm As}/{\rm In}_{0.52}^{}{\rm Al}_{0.48}^{}{\rm As}$--type heterostructure. The slope
of the fitted straight line and the point of its intersection with the ordinate axis yield
the values $\Omega=2.47\,{\rm meV}$ and $g=4.25$ for the spin precession frequency in the
absence of magnetic field and the $g$--factor, respectively. This values are in good
agreement with the results $\Omega=2.46\,{\rm meV}$ and $g=4.4\pm 0.2$ obtained in
Ref.~\onlinecite{das_etal}.

It should be stressed that $\Omega_{B}^{}=\Omega\approx 2k_{F}^{}\alpha$ only at $g=2$. Only
in this case the beating nodes locations are strictly periodic in the reverse magnetic field
and are described by the condition $2k_{F}^{}\alpha=(k+1/2)\omega_{c}^{}$ obtained in
Refs.~\onlinecite{taras_etal_2002,averk_etal_2005}.

\begin{figure}[t!]
\begin{center}
\includegraphics[scale=0.56]{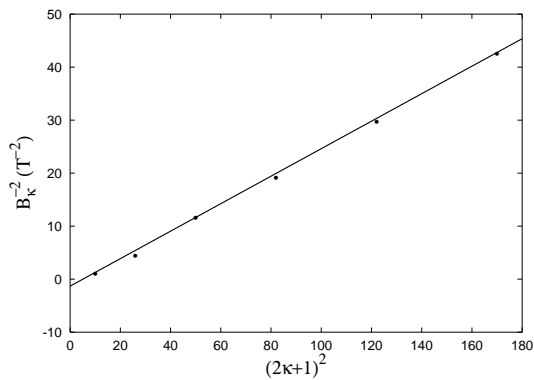}
\caption {Measured locations, $B_{k}^{}$, of the SdH oscillation beating
nodes as compared to the behavior predicted by Eq.~(\ref{eq:node}). The
points correspond to the values $B_{1}^{}=0.873\,{\rm T}$, $B_{2}^{}=0.460\,
{\rm T}$, $B_{3}^{}=0.291\,{\rm T}$, $B_{4}^{}=0.227\,{\rm T}$, $B_{5}^{}=
0.183\,{\rm T}$, $B_{6}^{}=0.153\,{\rm T}$, measured for the
${\rm In}_{0.65}^{}{\rm Ga}_{0.35}^{}{\rm As}/{\rm In}_{0.52}^{}
{\rm Al}_{0.48}^{}{\rm As}$--heterostructure \cite{das_etal}. The straight
line is the result of least--squares fitting of Eq.~(\ref{eq:node}).}
\end{center}
\end{figure}

Using Eq.~(\ref{eq:total_sigma}), we have performed numeric analysis of the
SdH oscillations of the longitudinal resistance $\rho$ (Fig.~5) and the Hall
coefficient $R_{H}^{}$ (Fig.~6) for parameters ($g=4,\,E_{0}=108.93\,{\rm meV},
\,\Omega =5.13\,{\rm meV}$, and $g=3.8,\,E_{0}=98.85\,{\rm meV},
\,\Omega =5.59\,{\rm meV}$) corresponding to the gate voltages
$V_{g}^{}=0\,{\rm V}$ and $V_{g}^{}=-0.3\,{\rm V}$ for the
${\rm In\,Ga\,As}/{\rm In\,Al\,As}$--heterostructure investigated in
Ref.~\onlinecite{nitta_etal}. The parameters $g$ and $\Omega $ were calculated 
from the positions of two successive nodes $B_{1}$, and $B_{2}$, using 
Eq.~(\ref{eq:node}), while $E_{0}$ was adjusted to the SdH oscillation period. 
At $V_{g}=0\,{\rm V}$, the calculated value $\Omega =5.13\,{\rm meV}$ is 
close to the result $\Omega =5.4\,{\rm meV}$ of Ref.~18, whereas at 
$V_{g}=-0.3\,{\rm V}$, the calculated value $\Omega =5.59\,{\rm meV}$ is in 
excellent agreement with the value  $\Omega =5.6\,{\rm meV}$ obtaned in 
Ref.~\onlinecite{nitta_etal}. The results of $\rho$ calculation are compared 
in Fig.~5 to the experimental curves from Ref.~\onlinecite{nitta_etal}. It can 
be seen that the theoretical results reproduce well the period and beating nodes 
location of the measured magnetoresistance oscillations. Some difference in 
oscillation amplitude is due to the fact that the temperature smearing of the 
Fermi level was not taking into account in our analysis for simplicity. 
The negative magnetoresistance observed in Ref.~\onlinecite{nitta_etal} 
lies outside the ladder approximation.

\begin{figure}[t!]
\vspace*{0.5cm}
\begin{center}
\includegraphics[scale=1.1]{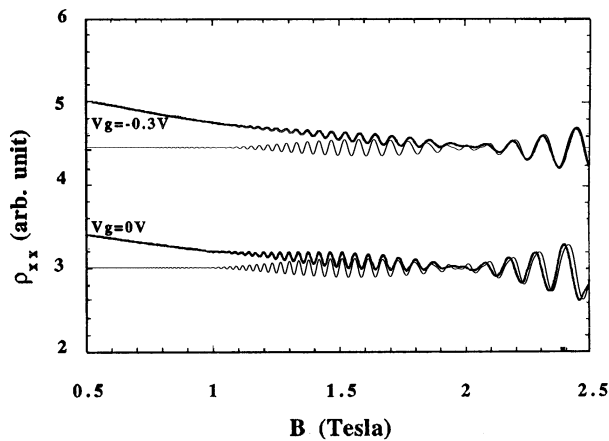}
\caption {The theoretical curves of the $\rho$ magnetooscillation as compared with
the measured \cite{nitta_etal} results for the ${\rm In\,Ga\,As}/{\rm In\,Al\,
As}$--heterostructure at gate voltages $V_{g}^{}=0\,{\rm V}$ and 
$V_{g}^{}=-0.3\,{\rm V}$. The experimental data are denoted by the thick solid line and 
the theoretical ones by thin solid line.}
\end{center}
\end{figure}

\begin{figure}[t!]
\begin{center}
\includegraphics[scale=0.56]{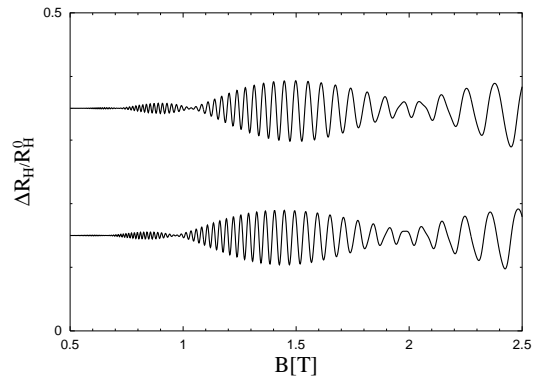}
\caption {Plots of the SdH oscillations of the Hall coefficient of the 2D Rashba
system calculated for the parameters corresponding to the ${\rm In\,Ga\,As}/
{\rm In\,Al\,As}$--heterostructure \cite{nitta_etal} for gate voltages
$V_{g}^{}=0\,{\rm V}$ and $V_{g}^{}=-0.3\,{\rm V}$ (from the bottom upwards).}
\end{center}
\end{figure}

In conclusion it should be stressed once more that the results of this work have
been obtained with rigorous account taken of both the spin--orbit and Zeeman
splitting of the energy levels. Up till now this was made by using numeric analysis
only \cite{wang_etal_2003,wang_etal_2005,yang_chang_2006}. Our results reproduce
quantitatively all the peculiarities of the magnetooscillation curves obtained in
such a way for models with Rashba spin--orbit interaction
\cite{wang_etal_2003,yang_chang_2006}, except for the anomalously large positive
magnetoresistance obtained in Ref.~\onlinecite{wang_etal_2003}. Our conclusion about the
absence of positive magnetoresistance is in drastic contradiction with
Ref.~\onlinecite{wang_etal_2003}, but agrees with the results of analytical
\cite{taras_etal_2002,averk_etal_2005} and recent numerical \cite{yang_chang_2006}
studies.

\acknowledgments

We thank A.K.~Arzhnikov, A.V.~Germanenko, G.I.~Kharus, G.M.~Minkov
and V.I.~Okulov for helpful discussions of results of this work.

This work was supported by the RFBR, grant 04--02--16614

\end{document}